\documentclass[a4paper,10pt,eqno]{article}
\usepackage{theorem}
\usepackage{latexsym,amssymb,amsfonts,amsmath}
\usepackage{graphicx}

\newtheorem{thm}{Theorem}[section]

\newtheorem{cor}[thm]{Corollary}
\newtheorem{exa}[thm]{Example}

\theoremheaderfont{\scshape}

\title{{\Large {\bf STATIONARY MEASURES OF THREE-STATE QUANTUM WALKS ON THE ONE-DIMENSIONAL LATTICE}}}
\author{
{\small Hikari Kawai\footnote{kawai-hikari-dy@ynu.jp},\quad Takashi Komatsu\footnote{komatsu-takashi-fn@ynu.ac.jp (e-mail of the corresponding author)},\quad Norio Konno\footnote{konno-norio-bt@ynu.ac.jp}}\\
{\scriptsize  Department of Applied Mathematics, Faculty of Engineering, Yokohama National University}\\
{\scriptsize \footnotesize\it 79-5 Tokiwadai, Hodogaya, Yokohama, 240-8501, Japan}\\
}
\date{\empty}
\begin{document}
\maketitle

\par\noindent
\begin{small}
\par\noindent
{\bf Abstract}. In this paper, we consider stationary measures of discrete-time three-state quantum walks including the Fourier and Grover walks in the one-dimensional lattice. We give non-uniform stationary measures by solving the corresponding eigenvalue problem. Our new method is based on a reduced matrix, which is different from the generating function approach in our previous work. As a corollary, the Fourier walk on the cycle has a stationary measure with a periodicity.

\footnote[0]{
{\it Abbr. title:} Limiting distributions of quantum walks on the square lattice
}
\footnote[0]{
{\it AMS 2000 subject classifications: }
60F05, 81P68
}
\footnote[0]{
{\it Keywords: } 
discrete-time quantum walk, stationary measure, periodicity
}
\end{small}

\setcounter{equation}{0}

\section{Introduction \label{intro}}
The notion of discrete-time quantum walks was introduced by Aharonov et al. \cite{adz} as a quantum analog of the classical one-dimensional random walks. It is known that the long-time asymptotic behavior of the transition probability for quantum walks on the one-dimensional lattice is quite different from that of classical random walks \cite{ko1}. Recently, the quantum walk is intensively studied in quantum physics and quantum computing \cite{mw}, \cite{por}.

One of the basic interests for quantum walks is to determine stationary measures of quantum walks. The stationary measures of Markov chains have been intensively investigated, however, the corresponding study of quantum walks has not been done enoughly. In 2013, Konno et al. \cite{kls} treated two-state quantum walks with one defect at the origin and showed that a stationary measure with exponential decay with respect to the location for the quantum walk starting from infinite sites is identical to a time-averaged limit measure for the same quantum walk starting from just the origin. Endo et al. \cite{ekst} got a stationary measure of the quantum walk with one defect whose coin matrices are defined by the Hadamard matrix. Endo and Konno \cite{ek} calculated a stationary measure of quantum walk with one defect which was introduced by Wojcik et al. \cite{wlkggb}. The stationary measure of the two-phase quantum walk with one defect and without defect was obtained by Endo et al. \cite{eekst} and Endo et al. \cite{eko}, respectively.

Konno and Takei \cite{kt} showed that the set of uniform measures is contained the set of stationary measures and gave non-uniform stationary measures. Konno \cite{ko2} obtained stationary measures of the three-state Grover walk. Wang et al. \cite{wlw} investigated stationary measures of the three-state Grover walk with one defect at the origin. Furthermore, Endo et al. \cite{ekk2} clarified a relation between stationary and limit measures of the three-state Grover walk. Endo et al. \cite{ekk1} obtained stationary measures for the diagonal quantum walks with one defect or without defect including the three-state model.

This paper is organized as follows. Section \ref{quantum3} is devoted to the definition of three-state discrete-time quantum walks on the one-dimensional integer lattice. In Section \ref{method}, we introduce a new method based on a reduced matrix, which is different from the generating function approach in our previous work. In Sections \ref{type1} and \ref{type2}, we obtain stationary measures of Types $1$ and $2$ by solving the eigenvalue problem, respectively. Moreover, we give their typical examples. Conclusions are given in  Section \ref{conclusion}.  
\section{Three-state discrete-time quantum walks}\label{quantum3}
In this section, we give the definition of three-state quantum walk on $\mathbb{Z}$, where $\mathbb{Z}$ is the set of integers. It is well known that the long-time asymptotic behavior of the transition probability for the three-state Grover walk on $\mathbb{Z}$ shows $localization$ \cite{iks}. This phenomenon is one of typical properties for discrete-time quantum walks \cite{ikk}, \cite{iks}, \cite{tate}, \cite{wkkk} which is not seen for usual classical random walks.

The discrete-time quantum walk on $\mathbb{Z}$ defined by a unitary matrix;
\begin{align*}
A=\begin{bmatrix}
a_{11}&a_{12}&a_{13}\\
a_{21}&a_{22}&a_{23}\\
a_{31}&a_{32}&a_{33}
\end{bmatrix}.
\end{align*}
We call this unitary matrix the coin matrix. To consider the time evolution, decompose the matrix $A$ as
\[A=P+R+Q\]
with
\[P=\begin{bmatrix}
a_{11}&a_{12}&a_{13}\\
0&0&0\\
0&0&0\end{bmatrix},\quad R=\begin{bmatrix}
0&0&0\\
a_{21}&a_{22}&a_{23}\\
0&0&0\end{bmatrix},\quad Q=\begin{bmatrix}
0&0&0\\
0&0&0\\
a_{31}&a_{32}&a_{33}
\end{bmatrix}.\]
A particle in the classical random walk moves at each step either one unit to the right with probability $p$ or one unit to the left with probability $q$, where $p+q=1$, $p$, $q>0$. On the other hand, the discrete-time quantum walk describes not only the motion of a particle but also the change of the states of a particle. Let $\mathbb{C}$ be the set of complex numbers. The state at time $n$ and location $x$ can be expressed by a three-dimensional vector:
\[\Psi_{n}(x)=\begin{bmatrix}\Psi^{L}_{n}(x)\\ \Psi^{O}_{n}(x)\\ \Psi^{R}_{n}(x) \end{bmatrix}\in\mathbb{C}^{3}\quad (x\in\mathbb{Z}, \ n\in\mathbb{Z}_{\geq}),\]
where $\mathbb{Z}_{\geq}=\{0,1,2,\ldots\}$. The time evolution of a quantum walk with a coin matrix $A$ is defined by the unitary operator $U_A$ in the following way:
\begin{equation*}
\Psi_{n+1}(x)\equiv(U_A\Psi_{n})(x)=P\Psi_{n}(x+1)+R\Psi_{n}(x)+Q\Psi_{n}(x-1).
\end{equation*}
 This equation means that the particle moves at each step one unit to the right with matrix $P$ or one unit to the left with matrix $Q$. The particle stays at the current location with matrix $R$. For time $n\in\mathbb{Z}_{\geq}$ and location $x\in\mathbb{Z}$, we define the measure $\mu_n(x)$ by
$$\mu_n(x)=\|\Psi_n(x)\|_{\mathbb{C}^3}^2,$$
where $\|\cdot\|_{\mathbb{C}^3}$ denotes the standard norm on $\mathbb{C}^3$. Let $\mathbb{R}_{\geq}=[0,\infty)$.  Here we introduce a map $\phi:(\mathbb{C}^3)^{\mathbb{Z}}\longrightarrow (\mathbb{R}_{\geq})^{\mathbb{Z}}$ such that if
$$\Psi_n={}^T\begin{bmatrix}\cdots, \begin{bmatrix}\Psi^{L}_{n}(-1)\\ \Psi^{O}_{n}(-1)\\ \Psi^{R}_{n}(-1)\end{bmatrix}, \begin{bmatrix}\Psi^{L}_{n}(0)\\ \Psi^{O}_{n}(0)\\ \Psi^{R}_{n}(0)\end{bmatrix}, \begin{bmatrix}\Psi^{L}_{n}(1)\\ \Psi^{O}_{n}(1)\\ \Psi^{R}_{n}(1)\end{bmatrix},\cdots\end{bmatrix}\in(\mathbb{C}^{3})^{\mathbb{Z}},$$
then 
$$\phi(\Psi_n)={}^T\begin{bmatrix}\cdots, \sum_{j=L}^{R}|\Psi_n^{j}(-1)|^2, \sum_{j=L}^{R}|\Psi_n^{j}(0)|^2, \sum_{j=L}^{R}|\Psi_n^{j}(1)|^2,\cdots\end{bmatrix}\in(\mathbb{R}_{\geq})^{\mathbb{Z}}.$$
Thus for any $x\in\mathbb{Z}$, we get
$$\phi(\Psi_n)(x)=\psi(\Psi_n(x))=\sum_{j=L}^{R}|\Psi_n^{j}(x)|^2=\mu_n(x).$$

\section{Stationary measure and our method}\label{method}
\subsection{Definition of stationary measure for quantum walk}
\noindent
In this subsection, we give the definition of stationary measure for the quantum walk. We define a set of measures, $\mathcal{M}_s(U_A)$, by
\begin{equation*}
\begin{split}
&\mathcal{M}_s(U_A)=\Big\{\mu\in[0,\infty)^\mathbb{Z}\setminus\{\textbf{0}\};\ there\ exists\ \Psi_0\in(\mathbb{C}^3)^{\mathbb{Z}}\ such\ that\\
&\hspace{8.0cm}\phi\big(U_A^n\Psi_0\big)=\mu\ \big(n=0,1,2, \ldots\big) \Big\}.
\end{split}
\end{equation*}
where $\textbf{0}$ is the zero vector. Here $U_A$ is time evolution operator of quantum walk associated with a unitary matrix $A$. We call this measure $\mu\in\mathcal{M}_s(U_A)$ the stationary measure for the quantum walk defined by the unitary operator $U_A$. If $\mu\in\mathcal{M}_s(U_A)$, then $\mu_n=\mu$ for $n\in\mathbb{Z}_{\geq}$, where $\mu_n$ is the measure of quantum walk given by $U_A$ at time $n$.

Next we consider the following eigenvalue problem of the quantum walk determined by $U_A$:
\begin{equation}\label{eig.pro}
U_A\Psi=\lambda\Psi\quad(\lambda\in\mathbb{C},\ |\lambda|=1).
\end{equation}
Then we see that $\phi(\Psi)\in\mathcal{M}_s(U_A)$. Our purpose of this paper is to find stationary measures for our three-state quantum walks by using Eq. \eqref{eig.pro}.
\subsection{Our method}
In this subsection, we introduce our new method to obtain the stationary measures for the three-state quantum walks. First, we see that $U_A\Psi=\lambda\Psi$ is equivalent to the following relations:
\begin{equation}\label{equation1}
 \begin{cases}
      \lambda\Psi^L(x)=a_{11}\Psi^L(x+1)+a_{12}\Psi^{O}(x+1)+a_{13}\Psi^{R}(x+1) ,                                                                                
      \\
      \lambda\Psi^O(x)=a_{21}\Psi^L(x)+a_{22}\Psi^{O}(x)+a_{23}\Psi^{R}(x),
    \\
     \lambda\Psi^R(x)=a_{31}\Psi^L(x-1)+a_{32}\Psi^{O}(x-1)+a_{33}\Psi^{R}(x-1) .                                                                                   
      \end{cases}
\end{equation}
Then we can rewrite Eq. $(\ref{equation1})$ as
\begin{equation}
 \begin{cases}\label{equation2}
      \Psi^O(x)=\frac{1}{\lambda-a_{22}}\Big\{a_{21}\Psi^L(x)+a_{23}\Psi^R(x)\Big\} ,                                                                                
      \\
      \lambda\Psi^L(x)=\Big(a_{11}+\frac{a_{12}a_{21}}{\lambda-a_{22}}\Big)\Psi^L(x+1)+\Big(a_{13}+\frac{a_{12}a_{23}}{\lambda-a_{22}}\Big)\Psi^{R}(x+1),
    \\
     \lambda\Psi^R(x)=\Big(a_{31}+\frac{a_{21}a_{32}}{\lambda-a_{22}}\Big)\Psi^L(x-1)+\Big(a_{33}+\frac{a_{23}a_{32}}{\lambda-a_{22}}\Big)\Psi^{R}(x-1).                                                                                   
      \end{cases}
\end{equation}
Since $\Psi^{O}(x)$ is expressed by $\Psi^{L}(x)$ and $\Psi^{R}(x)$ from Eq. \eqref{equation2}, we consider only $\Psi^{L}(x)$ and $\Psi^{R}(x)$. Thus we get
\begin{multline}\label{equation6}
\lambda\begin{bmatrix}\Psi^{L}(x)\\ \Psi^{R}(x) \end{bmatrix}=
\begin{bmatrix}
a_{11}+\frac{a_{12}a_{21}}{\lambda-a_{22}}&a_{13}+\frac{a_{12}a_{23}}{\lambda-a_{22}}\\
0&0
\end{bmatrix}
\begin{bmatrix}\Psi^{L}(x+1)\\ \Psi^{R}(x+1) \end{bmatrix}\\
+\begin{bmatrix}
0&0\\
a_{31}+\frac{a_{21}a_{32}}{\lambda-a_{22}}&a_{33}+\frac{a_{23}a_{32}}{\lambda-a_{22}}
\end{bmatrix}
\begin{bmatrix}\Psi^{L}(x-1)\\ \Psi^{R}(x-1) \end{bmatrix}.
\end{multline}
\noindent Here we introduce a $2\times2$ reduced matrix $A^{(Re)}$ derived from our $3\times3$ coin matrix $A$ as follows:
\begin{align}\label{redmat}
A^{(Re)}=\frac{1}{\lambda-a_{22}}\begin{bmatrix}
\lambda a_{11}-B&\lambda a_{13}+C\\
\lambda a_{31}+D&\lambda a_{33}-E
\end{bmatrix},
\end{align}
where we put
\begin{multline*}
B=\det\left(\begin{bmatrix}
a_{11}&a_{12}\\
a_{21}&a_{22}
\end{bmatrix}\right), \quad 
C=\det\left(\begin{bmatrix}
a_{12}&a_{13}\\
a_{22}&a_{23}
\end{bmatrix}\right),\\
D=\det\left(\begin{bmatrix}
a_{21}&a_{22}\\
a_{31}&a_{32}
\end{bmatrix}\right),\quad
E=\det\left(\begin{bmatrix}
a_{22}&a_{23}\\
a_{32}&a_{33}
\end{bmatrix}\right).
\end{multline*}
\noindent To obtain stationary measures of the quantum walk given by the coin matrix $A$, we focus on the reduced matrix $A^{(Re)}$ given by Eq. \eqref{redmat}. 

From now on, we treat the two classes of three-state quantum walks, i.e., Type $1$ and Type $2$. Then we suppose that 
\begin{equation*}
a_{ij}\ne0 \quad (1\leq i, j\leq3), \qquad|a_{22}|\ne1.
\end{equation*}
\begin{itemize}
  \item Type $1$ : We assume that $\lambda=-\frac{C}{a_{13}}=-\frac{D}{a_{31}}$ with $|\lambda|=1$.  Then a reduced matrix $A^{(Re)}$ is  
\begin{equation}
A^{(Re)}=\begin{bmatrix}
\tilde{a}_1&0\\
0&\tilde{a}_2
\end{bmatrix},
\end{equation}
where 
$$\tilde{a}_1=a_{11}-\frac{a_{13}a_{21}}{a_{23}},\quad \tilde{a}_2=a_{33}-\frac{a_{23}a_{31}}{a_{21}}.$$
  \item Type $2$ : We assume that $\lambda=\frac{B}{a_{11}}=\frac{E}{a_{33}}$ with $|\lambda|=1$. Then a reduced matrix $A^{(Re)}$ is  
\begin{equation*}
A^{(Re)}=\begin{bmatrix}
0&\tilde{a}_1\\
\tilde{a}_2&0
\end{bmatrix},
\end{equation*}
where 
$$\tilde{a}_1=a_{13}-\frac{a_{11}a_{23}}{a_{21}},\quad \tilde{a}_2=a_{31}-\frac{a_{21}a_{33}}{a_{23}}.$$
 \end{itemize}
\section{Result of Type 1}\label{type1}
\noindent In the previous section, we introduced a reduced matrix $A^{(Re)}$ to obtain stationary measures of quantum walks for Type $1$ and Type $2$. In this section, by using $A^{(Re)}$, we present a solution of eigenvalue problem, $U_A\Psi=\lambda\Psi$, for the three-state quantum walk of Type $1$. To do so, put $\Psi^L(x)\equiv\varphi_1$ and $\Psi^{R}(x)\equiv\varphi_3$ with $\varphi_1$, $\varphi_3\in\mathbb{C}$. Then, we have the following result which gives an explicit form of the solution $\Psi$ of the eigenvalue problem.
\begin{thm}\label{theorem1}
We assume that $\lambda=-\frac{C}{a_{13}}=-\frac{D}{a_{31}}$ with $|\lambda|=1$. Then we get
\begin{align*}
\Psi(x)=\begin{bmatrix}
(\tilde{a}_1^{-1}\lambda)^x\varphi_1\\
-\frac{a_{13}}{a_{12}a_{23}}\Big\{a_{21}(\tilde{a}_1^{-1}\lambda)^x\varphi_1+a_{23}(\tilde{a}_2\lambda^{-1})^x\varphi_3\Big\}\\
(\tilde{a}_2\lambda^{-1})^x\varphi_3\\
\end{bmatrix},
\end{align*}
where 
$$\tilde{a}_1=a_{11}-\frac{a_{13}a_{21}}{a_{23}},\quad \tilde{a}_2=a_{33}-\frac{a_{23}a_{31}}{a_{21}}.$$
\end{thm}
{\bf Proof.}
Suppose that $a_{11}\ne0$, $a_{33}\ne0$ and $|\lambda|=1$. From Eq. \eqref{equation6} and assumptions, $\lambda=-\frac{C}{a_{13}}=-\frac{D}{a_{31}}$ with $|\lambda|=1$, we get
\begin{equation*}
\lambda\begin{bmatrix}\Psi^{L}(x)\\ \Psi^{R}(x) \end{bmatrix}=
\begin{bmatrix}
\tilde{a}_1&0\\
0&0
\end{bmatrix}
\begin{bmatrix}\Psi^{L}(x+1)\\ \Psi^{R}(x+1) \end{bmatrix}
+\begin{bmatrix}
0&0\\
0&\tilde{a}_2
\end{bmatrix}
\begin{bmatrix}\Psi^{L}(x-1)\\ \Psi^{R}(x-1) \end{bmatrix}
=\begin{bmatrix}\tilde{a}_1\Psi^{L}(x+1)\\ \tilde{a}_2\Psi^{R}(x-1) \end{bmatrix}.
\end{equation*}
Then we have 
\begin{equation*}
\Psi^{L}(x+1)=(\tilde{a}_1^{-1}\lambda)\Psi^{L}(x),\quad \Psi^{R}(x)=(\tilde{a}_2\lambda^{-1})\Psi^{R}(x-1),
\end{equation*}
\begin{equation*}
\Psi^O(x)=\frac{1}{\lambda-a_{22}}\Big\{a_{21}\Psi^L(x)+a_{23}\Psi^R(x)\Big\}.
\end{equation*}
Therefore a solution $\Psi$ of $U_A\Psi=\lambda\Psi$ is given by
\begin{align*}
\Psi(x)=\begin{bmatrix}
(\tilde{a}_1^{-1}\lambda)^x\varphi_1\\
-\frac{a_{13}}{a_{12}a_{23}}\Big\{a_{21}(\tilde{a}_1^{-1}\lambda)^x\varphi_1+a_{23}(\tilde{a}_2\lambda^{-1})^x\varphi_3\Big\}\\
(\tilde{a}_2\lambda^{-1})^x\varphi_3\\
\end{bmatrix}\;\;\;(x\in \mathbb{Z}),
\end{align*}
where $\Psi^{L}(0)\equiv\varphi_1$, $\Psi^{R}(0)\equiv\varphi_3$ and $\varphi_1$, $\varphi_3\in\mathbb{C}^2$ with $|\varphi_1|+|\varphi_3|>0$. This completes the proof of Theorem \ref{theorem1}.
\vspace{0.3cm}

Next we present some examples by Theorem \ref{theorem1}.
\vspace{0.3cm}
\begin{exa}
\end{exa}

We consider the three-state Grover walk given by the $3\times3$ Grover matrix as
\begin{align}\label{grover111}
A_G=\frac{1}{3}
\begin{bmatrix}
-1&2&2\\
2&-1&2\\
2&2&-1
\end{bmatrix} .
\end{align}
Then we have $\lambda=-1$ and $\tilde{a}_1=\tilde{a}_2=-1$. Thus the reduced matrix becomes
\begin{align}\label{Gr1}
A_G^{(Re)}=
\begin{bmatrix}
-1&0\\
0&-1
\end{bmatrix}.
\end{align}
From Theorem \ref{theorem1} and Eq. \eqref{Gr1}, we get
\begin{align*}
\Psi(x)=
\begin{bmatrix}
\varphi_1\\
-(\varphi_1+\varphi_3)\\
\varphi_3
\end{bmatrix}\;\;\;(x\in\mathbb{Z}) .
\end{align*}
Let $\mathfrak{Re}(z)$ be the real part of complex number $z$. So we have a stationary measure of the Grover walk as
\begin{align*}
\mu(x)=2\{|\varphi_1|^2+|\varphi_3|^2+\mathfrak{Re}(\varphi_1\overline{\varphi_3})\}.
\end{align*}
In this example, the stationary measure is a uniform measure.
\begin{exa}\label{fourierex}
\end{exa}

We consider the three-state Fourier walk defined by the $3\times3$ Fourier matrix as
\begin{align}\label{fourier1}
A_F=\frac{1}{\sqrt{3}}
\begin{bmatrix}
1&1&1\\
1&\omega&\omega^2\\
1&\omega^2&\omega
\end{bmatrix},
\end{align}
where $\omega=e^{\frac{2\pi i}{3}}$.
 Then we have $\lambda=i$ and $\tilde{a}_1=e^{-\frac{\pi i}{6}}$ and $\tilde{a}_2=-i$. Therefore the reduced matrix becomes
\begin{align}\label{Fo1}
A_F^{(Re)}=
\begin{bmatrix}
e^{-\frac{\pi i}{6}}&0\\
0&i
\end{bmatrix}.
\end{align}
It follows from Theorem \ref{theorem1} and Eq. \eqref{Fo1} that
\begin{align}\label{fourieram1}
\Psi(x)=
\begin{bmatrix}
\omega^x\varphi_1\\
-(\omega^{x+1}\varphi_1+\varphi_3)\\
\varphi_3
\end{bmatrix}\;\;\;(x\in\mathbb{Z}).
\end{align}
Thus a stationary measure of the Fourier walk is given by
\begin{align*}
\mu(x)=2\{|\varphi_1|^2+|\varphi_3|^2+\mathfrak{Re}(\omega^{x+1}\varphi_1\overline{\varphi_3})\}.
\end{align*}
If we take $\varphi_1=\omega$, $\varphi_3=\omega^2$, then
\begin{align*}
\mu(x)=2\{2+\mathfrak{Re}(\omega^x)\}=\begin{cases}
      6 ,                                                                                &  \text{ $x=3m\quad (m\in\mathbb{Z}_{\geq}),$}\\
    3,
     &\text{$x=3m+1,\ 3m+2\quad (m\in\mathbb{Z}_{\geq}).$}
      \end{cases}
\end{align*}
Therefore the stationary measure of the Fourier walk is not the uniform measure. Furthermore,  the measure has period $3$. This is the first time for the Fourier walk that we found the stationary measure with a periodicity. Moreover, we can apply this example to the three-state Fourier walk on the cycle with $3m$ nodes $(m\in\mathbb{Z}_{>})$ and get the stationary measure with period $3$. More precisely, it is shown that the solution of the eigenvalue problem $U_A\Psi=\lambda\Psi$ for the three-state Fourier walk on $\mathbb{Z}$ given by Eq. \eqref{fourieram1} implies a solution of the eigenvalue problem on $C_{3m}$ $(m\in\mathbb{Z}_{>})$ with the following boundary conditions: 
\begin{equation*}
 \begin{cases}
     \sqrt{3}i\Psi^{R}(0)=\Psi^{L}(3m-1)+\omega^2\Psi^{O}(3m-1)+\omega\Psi^{R}(3m-1) ,                                                                                
      \\
    \sqrt{3}i\Psi^L(3m-1)=\Psi^{L}(0)+\omega^2\Psi^{O}(0)+\omega\Psi^{R}(0).
      \end{cases}
\end{equation*}
\noindent Here $C_N$ is the cycle where the number of the vertices is $N$. Therefore we have
\vspace{0.3cm}
\begin{cor}\label{cycle}
We consider the three-state Fourier walk defined by Eq. \eqref{fourier1} on a cycle with $3m$ $(m\in\mathbb{Z}_{\geq})$ nodes. Then the stationary measure of this quantum walk has the stationary measure with period $3$.
\end{cor}
\vspace{0.3cm}
\begin{exa}
\end{exa}

We consider a class of quantum walks determined by the $3\times3$ unitary matrices $A_1({\eta})$ $(\eta\in[0,2\pi))$ introduced by Stefanak et al. $\cite{sbj}$ as
\begin{align}\label{grover2}
A_1({\eta})=\frac{1}{6}
\begin{bmatrix}
-1-e^{2i\eta}&2(1+e^{2i\eta})&5-e^{2i\eta}\\
2(1+e^{2i\eta})&2(1-2e^{2i\eta})&2(1+e^{2i\eta})\\
5-e^{2i\eta}&2(1+e^{2i\eta})&-1-e^{2i\eta}
\end{bmatrix} ,
\end{align}
where $\eta\in[0,2\pi)$. Note that the quantum walk determined by $A_1({0})$ becomes the Grover walk. Then we have $\tilde{a}_1=\tilde{a}_2=-1$ and 
\begin{equation*}
\lambda=\frac{10-26\cos(2\eta)-24i\sin(2\eta)}{26-10\cos(2\eta)}\equiv e^{i\xi},\qquad \cos\xi=\frac{10-26\cos(2\eta)}{26-10\cos(2\eta)}.
\end{equation*}
So the reduced matrix becomes
\begin{equation*}
A_1({\eta})^{(Re)}=
\begin{bmatrix}
-1&0\\
0&-1
\end{bmatrix}.
\end{equation*}
From Theorem \ref{theorem1}, we obtain
\begin{align*}
\Psi(x)=
\begin{bmatrix}
(-\lambda)^x\varphi_1\\
-(1-\frac{3}{2}\tan\eta\cdot i)\big\{(-\lambda)^x\varphi_1+(-\overline{\lambda})^x\varphi_3\big\}\\
(-\overline{\lambda})^x\varphi_3
\end{bmatrix}\;\;\;(x\in\mathbb{Z}).
\end{align*}
Especially, we set $\varphi_1=\varphi_3$ and $T_x(\cos \xi)=\cos(x\xi)$, where $T_x(u)$ is the Chebyshev polynomial of the first kind. Then we have
\begin{align*}
|\Psi^{O}(x)|^2&=\Big|-\big(1-\frac{3}{2}\tan\eta\cdot i\big)(-1)^x(\lambda^x+\overline{\lambda}^x)\varphi_1\Big|^2\\
&=\Big|-(2-3\tan\eta\cdot i)(-1)^xT_x(\cos\xi)\varphi_1\Big|^2\\
&=(4+9\tan^2\eta)T_x^2(\cos\xi)|\varphi_1|^2,\\
\end{align*}
Therefore we get 
\begin{equation*}
\mu(x)=\Big\{2+(4+9\tan^2\eta)T_x^2(\cos\xi)\Big\}|\varphi_1|^2.
\end{equation*}
\newpage
\begin{exa}
\end{exa}

We consider a class of quantum walks determined by the $3\times3$ unitary matrices $A_2({\rho})$ $(\rho\in(0,1))$ introduced by Stefanak et al. $\cite{sbj}$ as
\begin{align}\label{grover3}
A_2({\rho})=
\begin{bmatrix}
-\rho^2&\rho\sqrt{2(1-\rho^2)}&1-\rho^2\\
\rho\sqrt{2(1-\rho^2)}&2\rho^2-1&\rho\sqrt{2(1-\rho^2)}\\
1-\rho^2&\rho\sqrt{2(1-\rho^2)}&-\rho^2
\end{bmatrix} ,
\end{align}
where $\rho\in(0,1)$. Remark that the quantum walk determined by the matrix $A_2({1/\sqrt{3}})$ becomes the Grover walk. 

Then we have $\lambda=-1$ and $\tilde{a}_1=\tilde{a}_2=-1$. Thus the reduced matrix becomes
\begin{equation*}
A_2({\rho})^{(Re)}=
\begin{bmatrix}
-1&0\\
0&-1
\end{bmatrix}.
\end{equation*}
So we have
\begin{align*}
\Psi(x)=
\begin{bmatrix}
\varphi_1\\
-\frac{\sqrt{1-\rho^2}}{\sqrt{2}\rho}(\varphi_1+\varphi_3)\\
\varphi_3
\end{bmatrix}\;\;\;(x\in\mathbb{Z}).
\end{align*}
Therefore we obtain
\begin{equation*}
\mu(x)=\frac{1+\rho^2}{2\rho^2}\Big(|\varphi_1|^2+\varphi_3|^2\Big)+\frac{1-\rho^2}{\rho^2}\mathfrak{Re}(\varphi_1\overline{\varphi_3}).
\end{equation*}
We see that this stationary measure is a uniform measure.
\section{Result of Type 2}\label{type2}
In the previous section, we obtained the stationary measure for the three-state quantum walk (Type $1$) determined by the reduced matrix $A^{(Re)}$ whose off-diagonal component is zero (i.e., diagonal matrix). This section deals with the stationary measure for the three-state quantum walk (Type $2$) given by the reduced matrix $A^{(Re)}$ whose diagonal component is zero. By using $A^{(Re)}$, we present a solution of $U_A\Psi=\lambda\Psi$, for the quantum walk of Type $2$ for $\Psi^{L}(x)=\varphi_x\in\mathbb{C}$ $(x\in\mathbb{Z})$ with $\varphi\not\equiv\textbf{0}$. Here $\varphi\equiv\textbf{0}$ means that $\varphi_x=0$ $(x\in\mathbb{Z})$. The following result for Type $2$ is a counterpart of Theorem \ref{theorem1} for Type $1$.
\begin{thm}\label{theorem2}
We assume that $\lambda=\frac{B}{a_{11}}=\frac{E}{a_{33}}$ with $|\lambda|=1$ and $\lambda^2=\tilde{a}_1\tilde{a}_2$. Let $\{\varphi_x\}_{x\in\mathbb{Z}}$ be a sequence of complex numbers except for $\varphi\equiv\textbf{0}$. Then we get
\begin{align*}
\Psi(x)=\begin{bmatrix}
\varphi_x\\
-\frac{a_{11}}{a_{12}a_{21}}\Big\{a_{21}\varphi_x+a_{23}(\tilde{a}_1^{-1}\lambda)\varphi_{x-1}\Big\}\\
(\tilde{a}_1^{-1}\lambda)\varphi_{x-1}\\
\end{bmatrix}\;\;\;(x\in \mathbb{Z}).
\end{align*}
where 
$$\tilde{a}_1=a_{13}-\frac{a_{11}a_{23}}{a_{21}},\quad \tilde{a}_2=a_{31}-\frac{a_{21}a_{33}}{a_{23}}.$$
\end{thm}
{\bf Proof.}
Suppose that $\lambda=\frac{B}{a_{11}}=\frac{E}{a_{33}}$ with $|\lambda|=1$. From Eq. \eqref{equation6} and assumptions $\lambda=\frac{B}{a_{11}}=\frac{E}{a_{33}}$ with $|\lambda|=1$, we get
\begin{equation*}
\lambda\begin{bmatrix}\Psi^{L}(x)\\ \Psi^{R}(x) \end{bmatrix}=
\begin{bmatrix}
0&\tilde{a}_1\\
0&0
\end{bmatrix}
\begin{bmatrix}\Psi^{L}(x+1)\\ \Psi^{R}(x+1) \end{bmatrix}
+\begin{bmatrix}
0&0\\
\tilde{a}_2&0
\end{bmatrix}
\begin{bmatrix}\Psi^{L}(x-1)\\ \Psi^{R}(x-1) \end{bmatrix}
=\begin{bmatrix}\tilde{a}_1\Psi^{R}(x+1)\\ \tilde{a}_2\Psi^{L}(x-1) \end{bmatrix}.
\end{equation*}
Then we have 
\begin{equation}\label{theorem2eq2}
\Psi^{L}(x)=\lambda^{-1}\tilde{a}_1\Psi^{R}(x+1),\quad \Psi^{R}(x)=\lambda^{-1}\tilde{a}_2\Psi^{L}(x-1).
\end{equation}
From Eq. \eqref{theorem2eq2}, we see that this quantum walk (Type 2) must satisfy the condition $\lambda^2=\tilde{a}_1\tilde{a}_2$. Let $\{\varphi_x\}_{x\in\mathbb{Z}}$ be a sequence of complex numbers except for $\varphi\equiv\textbf{0}$. Put $\Psi^{L}(x)=\varphi_x$. Therefore we obtain
\begin{align*}
\Psi(x)=\begin{bmatrix}
\varphi_x\\
-\frac{a_{11}}{a_{12}a_{21}}\Big\{a_{21}\varphi_x+a_{23}(\tilde{a}_1^{-1}\lambda)\varphi_{x-1}\Big\}\\
(\tilde{a}_1^{-1}\lambda)\varphi_{x-1}\\
\end{bmatrix}\;\;\;(x\in \mathbb{Z}),
\end{align*}
where 
$$\tilde{a}_1=a_{13}-\frac{a_{11}a_{23}}{a_{21}},\quad \tilde{a}_2=a_{31}-\frac{a_{21}a_{33}}{a_{23}}.$$
The proof of Theorem \ref{theorem2} is complete.
\vspace{0.1cm}
\begin{exa}
\end{exa}

We consider the Grover walk whose coin matrix is determined by the $3\times3$ unitary matrix $A_G$ given by the matrix $(\ref{grover111})$. 
Then we obtain $\lambda=\tilde{a}_1=\tilde{a}_2=1$. Thus the reduced matrix becomes
\begin{equation*}
A_G^{(Re)}=
\begin{bmatrix}
0&1\\
1&0
\end{bmatrix}.
\end{equation*}
Let $\{\varphi_x\}_{x\in\mathbb{Z}}$ be a sequence of complex numbers except for $\varphi\equiv\textbf{0}$. Then we get 
\begin{align*}
\Psi(x)=\begin{bmatrix}
\varphi_x\\
\frac{1}{2}\Big(\varphi_x+\varphi_{x-1}\Big)\\
\varphi_{x-1}\\
\end{bmatrix}\;\;\;(x\in \mathbb{Z}).
\end{align*}
Therefore we have
\begin{equation*}
\mu(x)=\frac{5}{4}\Big(|\varphi_x|^2+|\varphi_{x-1}|^2\Big)+\frac{1}{2}\mathfrak{Re}(\varphi_x\overline{\varphi_{x-1}}).
\end{equation*}
\vspace{0.1cm}
\begin{exa}\label{extype2,2}
\end{exa}

We consider a class of quantum walks determined by the $3\times3$ unitary matrix $A_1({\eta})$ given by the matrix $(\ref{grover2})$.
Then we have $\lambda=\tilde{a}_1=\tilde{a}_2=1$. So the reduced matrix becomes 
\begin{equation*}
A_1({\eta})^{(Re)}=
\begin{bmatrix}
0&1\\
1&0
\end{bmatrix}.
\end{equation*}
Let $\{\varphi_x\}_{x\in\mathbb{Z}}$ be a sequence of complex numbers except for $\varphi\equiv\textbf{0}$. Then we obtain
\begin{align*}
\Psi(x)=\begin{bmatrix}
\varphi_x\\
\frac{1}{2}\Big(\varphi_x+\varphi_{x-1}\Big)\\
\varphi_{x-1}\\
\end{bmatrix}\;\;\;(x\in \mathbb{Z}).
\end{align*}
Therefore we get
\begin{equation*}
\mu(x)=\frac{5}{4}\Big(|\varphi_x|^2+|\varphi_{x-1}|^2\Big)+\frac{1}{2}\mathfrak{Re}(\varphi_x\overline{\varphi_{x-1}}).
\end{equation*}
Interestingly, the stationary measure is independent of the parameter $\eta\in[0,2\pi)$.
\newpage
\begin{exa}
\end{exa}

We consider a class of quantum walks determined by the $3\times3$ unitary matrix $A_2({\rho})$ given by the matrix $(\ref{grover3})$. Then we have $\lambda=\tilde{a}_1=\tilde{a}_2=1$. That is, the reduced matrix becomes
\begin{equation*}
A_2({\rho})^{(Re)}=
\begin{bmatrix}
0&1\\
1&0
\end{bmatrix}.
\end{equation*}
Let $\{\varphi_x\}_{x\in\mathbb{Z}}$ be a sequence of complex numbers except for $\varphi\equiv\textbf{0}$. Then we get 
\begin{align*}
\Psi(x)=\begin{bmatrix}
\varphi_x\\
\frac{\rho}{\sqrt{2(1-\rho^2)}}\Big(\varphi_x+\varphi_{x-1}\Big)\\
\varphi_{x-1}\\
\end{bmatrix}\;\;\;(x\in \mathbb{Z}).
\end{align*}
Therefore we obtain
\begin{equation*}
\mu(x)=\frac{2-\rho^2}{2(1-\rho^2)}\Big(|\varphi_x|^2+|\varphi_{x-1}|^2\Big)+\frac{\rho^2}{1-\rho^2}\mathfrak{Re}(\varphi_x\overline{\varphi_{x-1}}).
\end{equation*}
In the previous example (Example \ref{extype2,2}), the stationary measure does not depend on the parameter $\eta\in[0,2\pi)$, but the stationary measure in this example depends on the parameter $\rho\in(0,1)$. 

In the rest of this section, we consider the $3\times3$ Fourier matrix defined by Eq. \eqref{fourier1}. Then we get
\begin{equation*}
\lambda=-\omega^2i,\quad \tilde{a}_1=\tilde{a}_2=-e^{\frac{\pi i}{6}}.
\end{equation*}
Thus we have $\lambda^2\ne\tilde{a}_1\tilde{a}_2$ which does not satisfy the assumption $\lambda^2=\tilde{a}_1\tilde{a}_2$ in Theorem \ref{theorem2}. So we do not apply Theorem \ref{theorem2} to the Fourier walk. In fact, we get
\begin{equation*}
\Psi^{L}(x)=\omega^2\Psi^{R}(x+1),\quad \Psi^{L}(x)=\omega\Psi^{R}(x+1).
\end{equation*}
Therefore the first equation is contradictory to the second one.
\section{Conclusion}\label{conclusion}
\noindent
In this paper, we obtained the stationary measures for the three-state quantum walks including the Fourier and Grover walks on $\mathbb{Z}$ by using the corresponding reduced matrix $A^{(Re)}$. As a special case, we found a stationary measure with periodicity. Moreover, this periodic stationary measure is also stationary measure on cycles. One of the future interesting problems would be to investigate the stationary measure for the general $N$-state quantum walk by using our new method introduced here. 
\par
\
\par
\noindent
{\bf Acknowledgements}
\noindent
This work is partially supported by the Grant-in-Aid for Scientific Research (Challenging Exploratory Research) of Japan Society for the Promotion of Science (Grant No.15K13443).

\par
\
\par

\begin{small}
\bibliographystyle{jplain}

\begin{thebibliography}{99}
\bibitem{adz}
Y. Aharonov, L. Davidovich and N. Zagury, {\it Quantum random walks}, Phys. Rev. A \textbf{48}, pp.1687-1690 (1993).
\bibitem{eekst}
S. Endo, T. Endo, N. Konno, E. Segawa and M. Takei, {\it Limit theorems of a two-phase quantum walk with one-defect}, Quantum Inf. Comput. \textbf{15}, pp.1373-1396 (2015).
\bibitem{ekk1}
T. Endo, H. Kawai and N. Konno, {\it The stationary measure for diagonal quantum walk with one defect}, arXiv:1603.08948 (2016).
\bibitem{ekk2}
T. Endo, H. Kawai and N. Konno, {\it Stationary measures for the three-state Grover walk with one defect in one dimension}, arXiv:1608.07402 (2016).
\bibitem{ek}
T. Endo and N. Konno, {\it The stationary measure of a space-inhomogeneous quantum walk on the line}, Yokohama Math. J., \textbf{60}, pp.33-47 (2014).
\bibitem{eko}
T. Endo, N. Konno and H. Obuse, {\it Relation between two-phase quantum walks and the topological invariant}, arXiv:1511.04230 (2015).
\bibitem{ekst}
T. Endo, N. Konno, E. Segawa and M. Takei, {\it A one-dimensional Hadamard walk with one defect}, Yokohama Math. J., \textbf{60}, pp.49-90 (2014).
\bibitem{ikk}
N. Inui, Y. Konishi and N. Konno, {\it Localization of two-dimensional quantum walks}, Phys. Rev. A \textbf{69}, 052323 (2004).
\bibitem{iks}
N. Inui, N. Konno and E. Segawa, {\it One-dimensional three-state quantum walk}, Phys. Rev. E \textbf{72}, 056112 (2005).
\bibitem{ko1}
N. Konno, {\it A new type of limit theorems for the one-dimensional quantum random walk}, J. Math. Soc. Japan \textbf{57}, pp.1179-1195 (2005).
\bibitem{ko2}
N. Konno, {\it The uniform measure for discrete-time quantum walks in one dimension}. Quantum Inf. Process.  \textbf{13}, pp.1103-1125 (2014).
\bibitem{kls}
N. Konno, T. Luczak and E. Segawa, {\it Limit measures of inhomogeneous discrete-time quantum walks in one dimension}, Quantum Inf. Process. \textbf{12}, pp.33-53 (2013).
\bibitem{kt}
N. Konno and M. Takei, {\it The non-uniform stationary measure for discrete-time quantum walks in one dimension}, Quantum Inf. Comput. \textbf{15}, pp.1060-1075 (2015).
\bibitem{mw}
K. Manouchehri and J. Wang, {\it Physical Implementation of Quantum Walks}, Springer (2013).
\bibitem{por}
R. Portugal, {\it Quantum Walks and Search Algorithms}, Springer (2013).
\bibitem{sbj}
M. Stefanak, I. Bezdekova and I. Jex, {\it Continuous deformations of the Grover walk preserving localization}, Eur. Phys. J. D \textbf{22}, 142 (2012).
\bibitem{tate}
T. Tate, {\it Eigenvalues, absolute continuity and localizations for periodic unitary transition operators}, arXiv:1411.4215 (2014).
\bibitem{wlw}
C. Wang, X. Lu and W. Wang, {\it The stationary measure of a space-inhomogeneous three-state quantum walk on then line}, Quantum Inf. Process. \textbf{14}, pp.867-880 (2015).
\bibitem{wkkk}
K. Watabe, N. Kobayashi, M. Katori and N. Konno, {\it Limit distributions of two-dimensional quantum walks}, Phys. Rev. A \textbf{77}, 062331 (2008).
\bibitem{wlkggb}
A. Wojcik, T. Luczak, P. Kurzynski, A. Grudka, T. Gdala and M. Bednarska-Bzdega, {\it Trapping a particle of a quantum walk on the line}, Phys. Rev. A \textbf{85}, 012329 (2012).
\end{thebibliography}

\end{small}

\end{document}